\documentstyle[prb,aps,epsfig,multicol]{revtex}

\begin{document}
\title{Electronic Structure and Fermiology of Sr$_3$Ru$_2$O$_7$}
\author{D.J. Singh and I.I. Mazin}
\address{Center for Computational Materials Science,\\
Naval Research Laboratory, Washington, DC 20375}
\date{\today}
\maketitle

\begin{abstract}
The electronic structure of layered Sr$_{3}$Ru$_{2}$O$_{7}$ in its
orthorhombic structure is investigated using density functional
calculations. The band structure near the Fermi energy, consists of Ru $%
t_{2g}$ states hybridized with O $p$ orbitals. The $yz$ and $xz$ bands,
which are largely responsible for the nesting related antiferromagnetic spin
fluctuations in Sr$_{2}$RuO$_{4}$ split pairwise into even and odd
combinations due to the interlayer coupling reducing the strength of
the nesting. The $xy$ bands show much less interplanar coupling as expected,
and also less $c$-axis dispersion, so that the barrel like sections are
largely intact compared to the single layer material. The zone folding due
to orthorhombicity yields small cylindrical lens shaped Fermi surfaces
centered at the midpoints of the former tetragonal $\Gamma -X$ lines. Fixed
spin moment calculations indicate that tetragonal Sr$_{3}$Ru$_{2}$O$_{7}$ is
borderline ferromagnetic but that orthorhombicity favors magnetism via a
substantial magnetoelastic coupling. These results are related to experimental
observations particularly in regard to magnetic properties.

\end{abstract}

\begin{multicols}{2}

\section{Introduction}

Perovskite derived ruthenates, (Sr,Ca)$_{n+1}$Ru$_{n}$O$_{3n+1}$ with
nominally Ru$^{4+}$ in an octahedral O environment display an
extraordinarily varied set of physical properties. In the Sr series, the end
members, SrRuO$_{3}$ and Sr$_{2}$RuO$_{4}$ are, respectively, an itinerant
ferromagnet \cite{randall,longo,kanbayashi,singh1,mazin1} and an
unconventional superconductor, widely thought to have triplet
pariing symmetry, \cite{maeno1,rice} possibly of magnetic origin.
\cite{mazin2,mazin3,kitaoka,mukuda}
The Fermi surface and momentum-dependent spin
fluctuations play key roles in the pairing in such scenarios for
superconductivity. The corresponding Ca end-points, CaRuO$_{3}$ and Ca$_{2}$%
RuO$_{4}$, which differ structurally from the corresponding Sr compounds
only by moderate lattice distortions, are a highly renormalized paramagnetic
metal and an antiferromagnetic Mott insulator, respectively. \cite%
{nakatsuji1,braden,puchkov1,cao1,cao2,cao2b,nakatsuji2} The multilayer
Ruddlesden-Popper compounds in the series show a wide variety of magnetic
orderings, metal insulator transitions and unusual transport properties,
with sample dependence. \cite{puchkov1,puchkov2,cao3,cao4,cao5,ikeda,cava}
Not surprisingly, strong magnetoelastic effects are found and have been
emphasized both theoretically \cite{mazin1} and experimentally. \cite%
{braden,nakatsuji2}

Density functional calculations of the band structure and Fermi surfaces of
superconducting Sr$_{2}$RuO$_{4}$ have been reported \cite{oguchi,singh2}
and subsequently confirmed in detail by quantum oscillation experiments. %
\cite{mackenzie,mackenzie2,bergemann} Angle resolved photoemission (ARPES)
experiments \cite{yokoya,lu} showed features very much like those in the
band calculations and quantum oscillation measurements but with small
shifts. However, because of van Hove singularities near the Fermi
energy, these shifts drastically
affect the topology. The Fermi surfaces predicted by band calculations
arise from three bands, one each from the three Ru $t_{2g}$
orbitals hybridized with the in-plane and, to a lesser extent, apical O $p$
states. The Fermi surface structure, which is highly two dimensional, thus
consists to zeroth order of planar sections perpendicular to $k_{x}$ and $%
k_{y}$, coming from the $yz$ and $xz$ orbitals, and a round cylinder centered
at $\Gamma $, derived from the $xy$ orbital. These hybridize and reconnect to
form three sections, a circular cylinder around $\Gamma $ and square
cylinders around $\Gamma $ and X. Strong nesting features remain from the $xz
$ and $yz$ derived sheets, which lead to soft spin-fluctuations around ${\bf %
k}=(2\pi/3a,2\pi/3a)$, as observed in neutron scattering. \cite{sidis} We 
have pointed out earlier\cite{mazin3} that the interplay of these
antiferromagnetic spin fluctuations and a broad Stoner background, related
to the proximity to ferromagnetism, may lead to competing order parameter
symmetries. However, the situation may be complicated by magnetoelastic
effects, that is, mixing between spin and charge degrees of freedom, {\it %
i.e.} spin-fluctuations and phonons, which could modify the pairing
interaction, perhaps favoring triplet symmetries though the physics of this
is not well understood.

Perhaps the most similar compound to Sr$_{2}$RuO$_{4}$ is the bilayer
compound Sr$_{3}$Ru$_{2}$O$_{7}$. \cite{muller,itoh} It is closely related
in structure, highly two dimensional and metallic. Cao and co-workers
reported ferromagnetism at 104 K with additional lower temperature
transitions, \cite{cao3} while other samples were not ferromagnetic at these
temperatures but exhibited transitions below 20 K. \cite{ikeda,cava}
Meanwhile, Huang and co-workers found no evidence for any long range
magnetic order down to 1.6 K \cite{huang} and Perry,
and Ikeda and co-workers report
that single crystals grown by a floating zone technique are paramagnetic,
with strongly enhanced susceptibility and high Wilson ratio, leading to
the conclusion that the material is
on the verge of ferromagnetism. \cite{perry,ik2}
Initially, Sr$_{3}$Ru$_{2}$O$_{7}$ was reported to be tetragonal,
$I4/mmm$, like Sr$_{2}$RuO$_{4}$.
\cite{muller} However, subsequent measurements revealed at least disordered
rotations of the octahedra. \cite{inoue,huang,shaked} Recently, Shaked and
co-workers \cite{shaked} have shown that the distortions are ordered and
refined the crystal structure into spacegroup $Bbcb$ with a rotation of
approximately 7 degrees. In contrast, superconducting Sr$_{2}$RuO$_{4}$ is
stable in the tetragonal $I4/mmm$ structure, but has a rotational phonon
mode that softens steeply from the $\Gamma $ point to the zone boundary
interacting with and crossing the acoustic branch ($\Sigma _{3}E_{u},$
corresponding at $(\pi ,\pi )$ to the main instability that produces the
orthorhombic structure in Sr$_{3}$Ru$_{2}$O$_{7}$). \cite{braden2}

The electronic structure of Sr$_{3}$Ru$_{2}$O$_{7}$ has been studied using
ARPES \cite{puchkov2} and density functional calculations \cite{hase} and
reasonable agreement was obtained, although not all the predicted sheets
were resolved experimentally. However, Perry and co-workers recently
reported low temperature Hall measurements on high quality ($\rho
_{res}\approx 2~\mu \Omega $~cm) single crystals that seem inconsistent with
the calculated tetragonal Fermi surfaces.\cite{perry}

The purpose of the present paper is to report density functional
calculations of the electronic structure, fermiology and magnetic properties
using the orthorhombic $Bbcb$ crystal structure.

\section{Structure and Method}

The calculations were done using the general potential linearized augmented
planewave (LAPW) method including local orbital extensions to treat the
upper core levels and relax linearization errors. \cite{singh-lapw,singh-lo}
The calculations were done within the local density approximation using the
Hedin-Lundqvist exchange correlation function. Spin polarized calculations
used the same functional and the von Barth-Hedin spin scaling. Well
converged basis sets and Brillouin zone samplings were used, similar to
those described previously. \cite{mazin2,mazin3,converg} Calculations for
the ideal tetragonal structure were based on the experimental structure of
Ref. \onlinecite{muller} which differs slightly from that used by Hase and
Nishihara. \cite{hase} However, the calculated electronic structure is
practically the same as obtained by them. The calculations for orthorhombic
Sr$_{3}$Ru$_{2}$O$_{7}$ are based on the neutron data of Shaked and
co-workers obtained on samples that, like floating-zone grown single crystals,
show a susceptibility peak around 20 K presumably due to magnetism
(but see below).
\cite{shaked}

\section{Band Structure and Fermiology}

\begin{figure}[tbp]
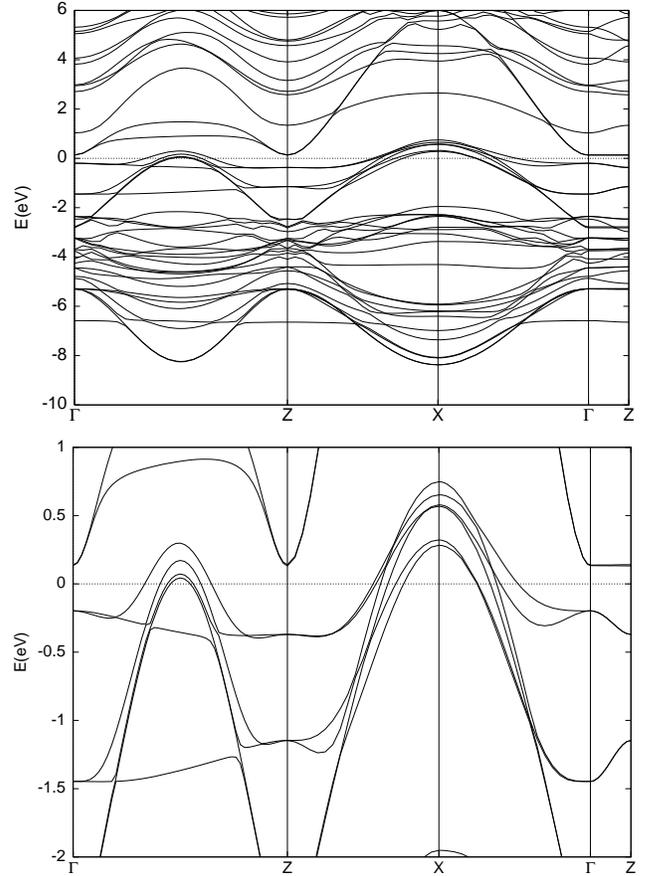

\centerline{\epsfig{file=band.epsi,angle=270,width=0.95\linewidth}}
\smallskip
\centerline{\epsfig{file=band-small.epsi,angle=270,width=0.95\linewidth}}
\setlength{\columnwidth}{3.2in} \nopagebreak
\smallskip
\caption{Band structure of Sr$_3$Ru$_2$O$_7$ in the ideal tetragonal
structure. The top panel is the valence band structure, and the bottom is a
blow-up around $E_F$.
The flat band at -6.6 eV is from a combination of Ru $d_{z^2}$ O $%
p_z$ bands that have little hopping to other sites in this structure.}
\label{bands}
\end{figure}

The calculated band structure and Fermi surfaces of tetragonal $I$4/mmm Sr$%
_{3}$Ru$_{2}$O$_{7}$ are shown in Figs. \ref{bands} and \ref{tet-fermi},
respectively. As may be seen, they are very similar to those reported
previously by Hase and Nishihara. \cite{hase} As mentioned, the three Fermi
surfaces of Sr$_{2}$RuO$_{4}$, which has one RuO$_{2}$ layer per cell, may
be regarded as arising from the three Ru $t_{2g}$ orbitals. The the $d_{xy}$
orbital gives rise to a round cylindrical electron-like sheet centered at $%
\Gamma $ (Z) and the $d_{xz}$ and $d_{yz}$ orbitals provide flat sheetlike
sections perpendicular to $k_{y}$ and $k_{z}$, respectively, that after
reconnection become square cylindrical sections around $X$ and $\Gamma $
along with strong nesting. To a first approximation, the Fermi surfaces of
tetragonal Sr$_{3}$Ru$_{2}$O$_{7}$ may be thought of as deriving from the
six same bands (three from each RuO$_{2}$ layer) with bonding -- antibonding
(odd -- even) splittings\cite{split} due to the interaction between the RuO$%
_{2}$ sheets comprising the bi-layer. However, as may be seen from the lack
of four-fold symmetry around the $X$ point in Fig. \ref{tet-fermi}, there is
more $k_{z}$ dispersion in tetragonal Sr$_{3}$Ru$_{2}$O$_{7}$ than in Sr$%
_{2} $RuO$_{4}$, particularly for the even bands.

There is an approximately square cylindrical hole pocket around $X$,
deriving from the odd combinations of $d_{xz}$ and $d_{yz}$ orbitals.
Another, cross-shaped, hole pocket  arises from reconnected sections of the
Fermi surfaces of the even and odd parity with the same orbital character.
The even combination also provides the innermost cylinder around $\Gamma $,
which like the other three $\Gamma $ centered cylinders is electron-like.
The second $\Gamma $ centered electron-like cylinder is also mainly $%
d_{xz}/d_{yz}$ derived near the (10) directions, but is mixed with the $%
d_{xy}$ character near (11).The two remaining $\Gamma/Z$ centered cylinders
are derived from the $d_{xy}$ orbitals, and have very little even-odd
splitting, in accord with their strongly in-plane orbital character. The
three outer $\Gamma $ centered cylinders change topology as the electron
count is increased as they touch along the basal plane $\Gamma -Z$ lines
giving rise to the van Hove singularities seen above $E_{F}$ in the density
of states, as in Sr$_{2}$RuO$_{4}$. It is noteworthy that even before
considering the effects of orthorhombicity, the Fermi surface nesting is
considerably reduced relative to Sr$_{2}$RuO$_{4}$. This is due to (i)
the even--odd splittings, (ii) the $k_{z}$ dispersions, and (iii), the
strong anticrossings of the {\it even} $d_{xz}/d_{yz}$ and the{\it \ odd} $%
d_{xy}$ bands. The striking difference from Sr$_{2}$RuO$_{4}$ stems from the
fact that nearest neighbor hopping between $d_{xz}/d_{yz}$ and $d_{xy}$
orbitals is forbidden within a single layer, but allowed across the layers
in a bi-layer. A $d_{xz}(d_{yz})$ orbital hybridizes with the $d_{xy}$ of the 
{\it opposite parity}, and this
is the strongest at $k=\{0,\pi/a\}$ ($\{\pi/a,0\}).$
Correspondingly, the strongest distortion of the $xz/yz$ band
occurs near the (1,1) directions. Note that additional hybridization between
the $xz/yz$ band and the $xy$ band (and the only one existing in Sr$_{2}$RuO$%
_{4})$ happens due to hopping across the SrO layer. Depending on $k_{z}$,
this hopping is of the same $(k_{z}=0)$ or of the opposite sign with respect
to the intra-bi-layer hopping. This brings about an additional $k_{z}$
dispersion.

Given the reduced nesting in Sr$_{3}$Ru$_{2}$O$_{7},$ one may conjecture
that the incommensurate antiferromagnetic spin-fluctuations seen in neutron
scattering experiments \cite{sidis} on Sr$_{2}$RuO$_{4}$ may be considerably
less prominent in Sr$_{3}$Ru$_{2}$O$_{7}$. On the other hand, tetragonal Sr$%
_{3}$Ru$_{2}$O$_{7}$ is much closer to a ferromagnetic instability. This is
not surprising considering the high value of the calculated density of
states (DOS) at the Fermi energy ($E_{F}$), $N(E_{F})$ = 4.5 states/eV Ru,
compared with 4.1 states/eV Ru for Sr$_{2}$RuO$_{4}$. \cite{singh2}

Non-spin-polarized local spin density approximation (LSDA) calculations for
the orthorhombic structure yield a still slightly larger $N(E_{F}$) = 5.0
states/eV Ru and the structure of the density of states around $E_{F}$ is
different (with larger magnitudes corresponding to slightly narrower $t_{2g}$
bands, as shown in Fig. \ref{dos}). The result is a magnetic ground state.
However, before turning to the magnetic properties, we discuss the
electronic properties of the non-spin-polarized system. As the magnetism has
substantially itinerant character (see below) the non-spin-polarized
electronic structure is expected to be a reasonable approximation for the
paramagnetic phase well above any ordering temperature.

As a reference point, we show in Fig. \ref{fold} the Fermi surfaces of the
tetragonal structure, but folded into the $Bbcb$ orthorhombic zone. This
zone is half the area of the tetragonal zone due to the cell doubling and is
rotated by 45$^{\circ }$. The most striking feature of the folding is the
development of cylindrical lens shaped Fermi surfaces centered at the
mid-points of the $\Gamma -Z$ lines (the mid-points of the $\Gamma -X$ and $%
Z-X$ lines in the tetragonal zone). These come from the two $d_{xy}$
circular cylinders around $\Gamma $ and the outer (odd) $d_{xz}/d_{yz}$
cylinder.

One may expect that such folded Fermi surfaces and particularly the lenses
should become observable in quantum oscillation experiments in samples with
ordered orthorhombic structures, and possibly in photoemission as shadow
Fermi surfaces with strength related to the matrix elements produced by the
orthorhombic distortion even in samples where the distortion is not fully
long range ordered, as long as the in-plane coherence length is large enough
(in analogy with some of the cuprate superconductors. \cite%
{bisco1,bisco2,bisco3,bisco4}) However, the actual distortion is a rotation
of approximately 7$^{\circ }$, which is not so small, and considerably
changes the Fermi surfaces beyond the naive zone folding picture.

In fact, significant differences between the folded tetragonal and
calculated orthorhombic Fermi surfaces are found, as shown in Fig. \ref%
{ort-fermi}. Hybridization gaps open where folded bands cross. As a result,
the square sections around the center of Fig. \ref{fold}, which came from
the overlap of the $d_{xy}$ derived bands, entirely disappear, while the
larger square section derived from the odd $d_{xz}/d_{yz}$ bands reconnects
with the the outer lens derived from the $d_{xy}$ bands. One can see 
remnants of this square in the dents near the tips of the outer lenses in
the orthorhombic structure. Also two small circular electron cylinders are
introduced around the $\Gamma -Z$ lines, with hardly any $z$ dispersion;
these are the bottoms of the $d_{x^{2}-y^{2}}$ bands, which in the
tetragonal structure were located a few mRy above the Fermi level (Fig.\ref%
{bands}). The lenses are now considerably distorted, because of the
above-mentioned interaction with the odd combination of $d_{xz}/d_{yz}$
states, and because of the increased the $k_{z}$ dispersion
(note the lack of reflection symmetry about the center line in Fig. \ref%
{ort-fermi}). Interestingly, this dispersion is concentrated mostly in small
and/or heavy sections, so the average (RMS) Fermi velocity remains quite two
dimensional, {\it i.e.} $v_{x}\approx v_{y}=1.3\times 10^{7}$ cm/s and $%
v_{z}=0.16\times 10^{7}$ cm/s. Within the constant scattering time
approximation this implies a resistivity anisotropy $\rho _{c}/\rho
_{a}\approx 70$ as compared to the LDA value for Sr$_{2}$RuO$_{4}$ of $%
\approx 300$, which is known to be an underestimate relative to experiment.
Clearly, the bilayer compound is much more three-dimensional than the single
layer one, and this, presumably, moves it still further from an
antiferromagnetic instability in-plane.

\begin{figure}[tbp]
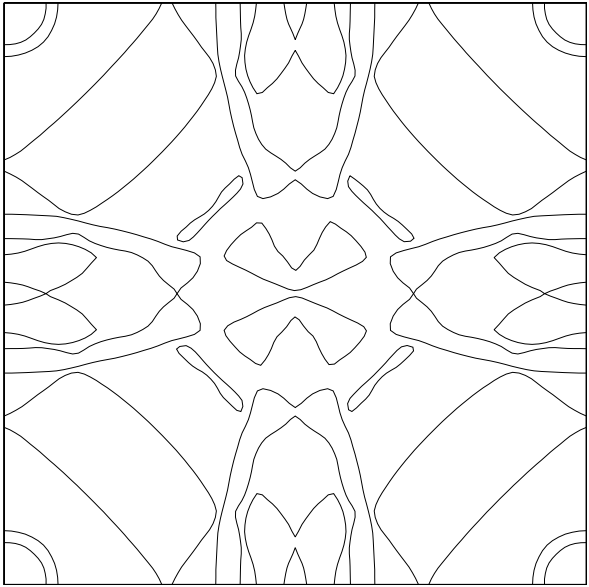

\centerline{\epsfig{file=850.epsi,angle=270,width=0.90\linewidth}}
\smallskip
\centerline{\epsfig{file=11.epsi,angle=270,width=0.90\linewidth}}
\setlength{\columnwidth}{3.2in} \nopagebreak
\smallskip
\caption{Basal plane (top) and mid-plane ($k_z$ shifted by 1/4, bottom)
Fermi surfaces of Sr$_3$Ru$_2$O$_7$ in the experimental
orthorhombic structure. Note that the zone is rotated 45 degrees with
respect to the tetragonal and folded.}
\label{ort-fermi}
\end{figure}

\begin{figure}[tbp]
\centerline{\epsfig{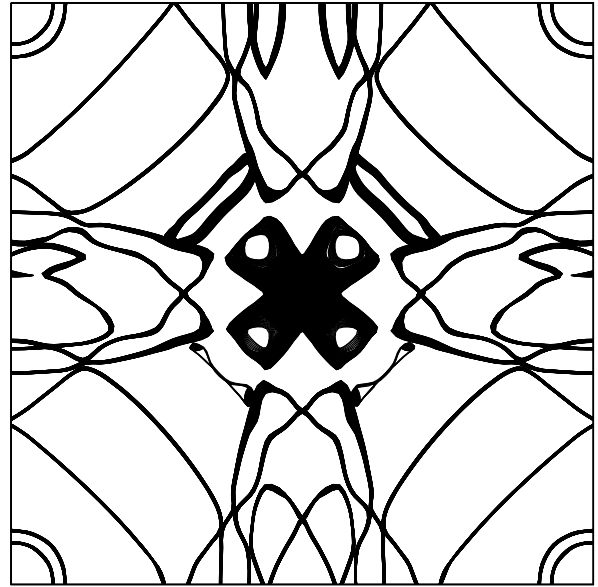}}
\setlength{\columnwidth}{3.2in} \nopagebreak
\smallskip
\caption{Fermi surfaces of orthorhombic Sr$_3$Ru$_2$O$_7$ as in
Fig. \ref{ort-fermi} but with filled areas indicating the effect of 
5 meV upwards and downwards shifts of $E_F$.}
\label{shift-fermi}
\end{figure}

Finally, it should be mentioned that, because of the hybridization gaps near
the center of the Brillouin zone of Fig. \ref{ort-fermi}, the Fermi surface
in this region is highly sensitive to the band filling. This is illustrated
in Fig. \ref{shift-fermi}, where Fermi surfaces calculated with 5 meV upwards
and downwards shifts of $E_{F}$ are shown, so that the thickness of the
lines is inversely proportional to the local Fermi velocity. Considering
the fact that LDA
calculations are sensitive to structural details and in any case should not
be relied on at the 5 meV level, the topology in this region cannot be
decided based on the present calculations. Because of this uncertainty, we
cannot unambiguously match our results with the Hall measurements of Perry
and co-workers \cite{perry}, but the possibility, suggested by them, that
orthorhombicity provides a way of reconciling their measurements with band
calculations seems plausible. If we restrict ourselves to the the ``stable''
sheets of the Fermi surface, we observe two electronic $d_{x^{2}-y^{2}}$
derived cylinders, the hole pseudo-square and the hole pseudo-cross formed
by the $d_{xz}/d_{yz}$ bands, and several sets of lenses, all electronic.
The complicated shape and changing curvature of these surfaces prevents us from
associating the volumes of these Fermi surface pockets with Hall numbers. The
only conclusion we can draw is that there is no clear way to guess the Hall
conductivity from the electronic structure of
the corresponding tetragonal compound. Furthermore, as
seen from Fig. \ref{shift-fermi}, shifting the chemical potential by as little
as 5 meV (corresponding to 55 K) drastically changes the topology of the
paramagnetic Fermi surfaces. The message of this observation is that
substantial and nontrivial temperature dependencies of the Hall coefficient
may occur independent of any anomalous Hall contribution.

\section{Magnetism}

As mentioned, in the idealized tetragonal structure, Sr$_{3}$Ru$_{2}$O$_{7}$
is found to be paramagnetic in the LSDA, but with a very high
susceptibility. In fixed spin moment calculations, moments up to 0.6 $\mu
_{B}$/Ru are induced with an energy cost of less than 4 meV/Ru -- a
number almost indistinguishable from zero considering the accuracy of the
present calculations. The orthorhombic distortion narrows the $t_{2g}$
manifold and increases the density of states leading to a magnetic ground
state via the Stoner mechanism. Furthermore, the paramagnetic orthorhombic
band structure shows extremely flat bands near the X point (cf. Fig. \ref%
{shift-fermi}), which create a low-weight but extremely sharp peak within a
few K from the Fermi level. Such a band structure is intrinsically unstable
against any symmetry lowering splitting of this peak, including a magnetic
instability. However, since the orthorhombic unit cell is already doubled,
an antiferromagnetic instability would work just as well. There is
nevertheless a strong reason for the system to prefer the ferromagnetic
coupling in plane. Generally speaking, the electronic structure of Ru based
perovskites shows strong Ru $d$ O $p$ hybridization and this favors
ferromagnetic couplings\cite{mazin1} (O $p$ states have rather large Hund
rule coupling). Within a local picture, this tendency may be viewed as
resulting from the fact that the density of states has substantial O weight
around $E_{F}$ and all things being equal, this favors magnetic
configurations where the O can polarize, {\it i.e.} ferromagnetism. Within a
simple Stoner picture, this O contribution to the magnetic energy is
proportional to the square of the O contribution to the density of states
around $E_{F}$. Decomposing $N(E_{F})$ onto the LAPW spheres, \cite{converg}
we find 56 \% Ru $d$, 20 \% plane O $p$, 2 \% apical (rocksalt) O $p$ and 3
\% bridging (interlayer) O $p,$ with the remainder divided between other
angular momentum characters (2 \%), and the interstitial (17 \%).
Considering the ionic radii, the interstitial contribution is likely more O $%
p$ than Ru $d$ derived, but in any case, there is substantial O $p$ 
involvement in $N(E_{F})$ and this contribution is dominated by the in-plane
oxygen. Not surprisingly then, ferromagnetic in-plane configurations are
favored. We find a self-consistent ferromagnetic (FM) solution with a spin
magnetization of 0.80 $\mu _{B}$/Ru and an energy of -23 meV/Ru relative
to the non-spin-polarized case. Calculations were also performed for
antiferromagnetic configurations with a (2 $\times $ 2) in-plane ordering,
and having adjacent Ru ions in the two planes of the bilayer polarized
parallel. However, no self-consistent magnetic configuration was found. Thus
it may be concluded that the magnetic character {\em within each plane} is
itinerant. However, calculations in which the Ru ions in a layer were
ferromagnetically aligned, but the layers were stacked antiferromagnetically
(so each bi-layer had one spin up and one spin down RuO$_{2}$ layer) did
yield a stable magnetic solution (denoted AF-A in the following), in this
case with an energy of -20 meV/Ru and a Ru moment (as measured by the
moment in a Ru sphere) only 14\% smaller than the ferromagnetic solution.

The energy difference between the FM and AF-A solutions contains two parts,
within the most simple model. The first is the interaction between the two
planes comprising the bi-layer. This, in turn, has a ferromagnetic part
originating from the Hund rule energy on the bridging
oxygen, and an antiferromagnetic
superexchange. Based on the calculation, we conclude that the former is
slightly stronger. In any case, it is likely larger than the inter bi-layer
coupling through the rock-salt layers. This expectation is based on the
geometry (hopping via two oxygens with unfavorable bond angles) and the fact
that the $k_{z}$ dispersion is considerably less than the anti-symmetric,
symmetric band splittings due to the interaction between the planes
comprising the bi-layer. Supposing that the inter-bilayer interaction is
anti-ferromagnetic (as expected for a superexchange coupling) and much
weaker than the intra bi-layer coupling, one may conclude that the latter is
ferromagnetic with a strength of order 3 meV/Ru. In this scenario, the
ground state is antiferromagnetic, consisting of ferromagnetic bi-layers,
stacked antiferromagnetically.

This represents a conceptually interesting
case for the Kosterlitz-Thouless theory of quasi-2D magnetic phase transitions.
\cite{KT} The difference from the textbook case is that the in-plane
magnetism is itinerant; the same qualitative picture applies, still, and so the
3D long range order (LRO)
transition temperature should be logarithmically suppressed compared to the
ferromagnetic-paramagnetic energy difference.
We can safely assume that the inter-bilayer coupling is less
than 30 K (the interplane coupling within a bilayer),
probably very much less, thus bringing about a substantial
Kosterlitz-Thouless suppression. One would, however, expect strong itinerant
spin fluctuations of ferromagnetic character in planes for temperatures well
above the LRO
transition temperature; these should manifest themselves, for instance,
in specific heat and magnetic susceptibility.
Very recently, Ikeda and co-workers reported a
detailed study of single crystal floating-zone Sr$_3$Ru$_2$O$_7$
including magnetic susceptibility, specific heat and resistivity plus
magnetic measurements under pressure. \cite{ik2} They report a 
susceptibility maximum at $T_{max} = 16$K accompanied by structure
in the resistivity, but
conclude that long range antiferromagnetic
order does not set in, based on the high nearly isotropic
$\chi(T)$ below $T_{max}$ and thus that the ground state is a
paramagnetic Fermi liquid on the verge of ferromagnetism.
In fact, the measured $\chi(T)$ is $\sim$ 15 times larger than
that of Sr$_2$RuO$_4$ (Ref. \onlinecite{m97}) and at least
as isotropic at low $T$.
Certainly, this is not expected with simple
local moment ordered magnetism, especially in a material with strong
magnetocrystalline anisotropy as is the case for magnetic ruthenates.
\cite{kanbayashi}
However, things are less clear cut in the itinerant metallic scenario
above. First of all, in the ordered itinerant case,
there is a Stoner continuum that contributes
to $\chi(T)$ along the direction of the moments (the low $\chi(T)$ direction);
normally this is a very small effect, but here the magnetocrystalline
anisotropy is expected to be very large \cite{kanbayashi} and the 
Fermi liquid very soft as evidenced by the strong Stoner renormalizations.
Secondly, it should be noted that as long as
the anisotropy is not in the trivial $c$-axis direction, the
twinning of orthorhombic samples will mean that there will be no orientation
where
the applied field is aligned with the moments, again presumably lowering
the anisotropy of $\chi$ in the magnetic state. However, both of these
mechanisms for lowering the anisotropy in $\chi$ would require fortuitous
strong numerical coincidences
to explain the observed isotropy of the measured $\chi$ in a magnetically
ordered state. On the other hand, the measured temperature dependence of
the two components of $\chi$ is notably different, which is not 
simply anticipated for a standard enhanced paramagnet either.
Another interesting possibility is to associate the
peak in $\chi(T)$ not with LRO antiferromagnetic ordering,
but with a Kosterlitz-Thouless type transition
associated with the 2D bilayers. In this case, there would be
no long range magnetic order below the peak, though the weak inter-bilayer
interactions could still yield a 3D ordered magnetic state at very low
temperatures. This scenario would also provide a convenient explanation
of the so called magnetic metallic phase in the Ca$_{2-x}$Sr$_x$RuO$_4$ phase
diagram between $x=0.2$ and $x=0.5$ where an unusual phase with a
peak in $\chi(T)$ is entered just as the system reaches an incipient
ferromagnetic instability with $x$ decreasing through 0.5. \cite{nakatsuji2}

Finally, under applied pressures of 1 GPa
Ikeda and co-workers \cite{ik2} report evidence for ferromagnetism though
with small moments ($M \approx 0.08 \mu_B$/Ru) starting at 70 K based
on magnetization measurements. While low moment ferromagnetism with this
Curie temperature cannot be excluded, such a high ratio of $T_C/M$ is
unusual in traditional materials.
Weak ferromagnetism due to canting seems to be more likely and also
fits better with our calculations.
We note that the ordering temperature in such a scenario would be sensitive
to the out-of-plane superexchange coupling, which in turn is expected to
increase strongly under pressure.

\section{Summary and Conclusions}

Density functional studies of Sr$_3$Ru$_2$O$_7$ show a substantial
coupling of the electronic and magnetic properties with the octahedral
rotations associated with the orthorhombic distortion. In particular, the
Fermi surface nesting (already lower than in Sr$_2$RuO$_4$) is further
reduced and the $t_{2g}$ manifold is narrowed leading to magnetic
tendencies. Using the experimental crystal structure \cite{shaked}
we find that the 2D RuO$_2$ planes form an itinerant system with
ferromagnetic ordering tendencies, and additionally that the coupling
between the two such planes comprising a bilayer is weak, but
also ferromagnetic. We argue that the $c$-axis coupling between bilayers
is weaker still and based on speculate that the magnetic character should
be highly two dimensional, perhaps exhibiting Kosterlitz-Thouless
physics. 

\acknowledgements

We are grateful for helpful discussions with C.S. Hellberg,
J.D. Jorgensen, J.W. Lynn, A.P.
Mackenzie, S. Nakatsuji and M. Sigrist. We thank J.D. Jorgensen for a
preprint of Ref. \onlinecite{shaked}. Computations were performed using
facilities of the DoD HPCMO ASC center. Work at the Naval Research Laboratory
is supported by the Office of the Naval Research.

\begin{figure}[tbp]
\centerline{\epsfig{file=tet-fermi.epsi,angle=270,width=0.90\linewidth}}
\setlength{\columnwidth}{3.2in} \nopagebreak
\smallskip
\caption{Basal plane Fermi surface of Sr$_3$Ru$_2$O$_7$ in the ideal
tetragonal structure. Note that because of interpolation in the plotting
program (even-odd) crossings that occur in this plane
become very small anti-crossings.}
\label{tet-fermi}
\end{figure}

\begin{figure}[tbp]
\centerline{\epsfig{file=dos-tet.epsi,angle=270,width=0.95\linewidth}}
\smallskip
\centerline{\epsfig{file=dos-orth.epsi,angle=270,width=0.95\linewidth}}
\setlength{\columnwidth}{3.2in} \nopagebreak
\smallskip
\caption{ Electronic density of states of ideal tetragonal (top) and
orthorhombic (bottom) Sr$_3$Ru$_2$O$_7$ on a per formula unit basis.
Energies are relative to $E_F$.}
\label{dos}
\end{figure}

\begin{figure}[tbp]
\centerline{\epsfig{file=fold.epsi,angle=0,width=0.90\linewidth}}
\setlength{\columnwidth}{3.2in} \nopagebreak
\smallskip
\caption{Basal plane Fermi surface of Sr$_3$Ru$_2$O$_7$ in the ideal
tetragonal structure folded into the orthorhombic zone (see text).}
\label{fold}
\end{figure}

\end{multicols}


\begin{references}
\bibitem{split} This may seem counterintuitive; however, the positions of the
Ru bands are mainly defined by the crystal field, which, in turn, is mainly
due to Ru-O hybridization. The odd combination of Ru d orbitals does not
hybridize with the bridging oxygen, and thus is less pushed up than the even
one.
\bibitem{converg} LAPW sphere radii of 2.1, 2.0 and 1.65 a$_0$ were used for
Sr, Ru and O, respectively. An LAPW sector basis set cut-off of 4.55 a$%
_0^{-1}$ was used with the addition of local orbitals for the Sr $s$ and $p$%
, Ru $s$, $p$ and $d$ and O $s$ channels.
\end{references}
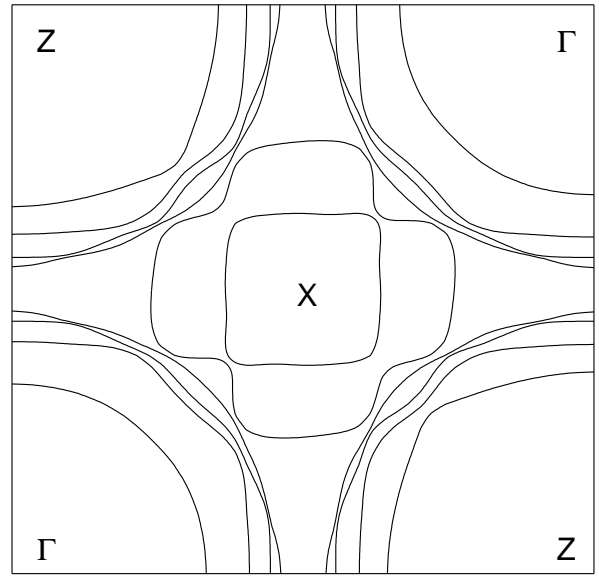
\end{document}